\documentclass[num-refs]{wiley-article}

\papertype{Original Article}
\paperfield{Journal Section}

\usepackage{epstopdf}
\usepackage[markup=underlined]{changes}
\usepackage{multirow}
\usepackage{subcaption}  

\title{Fully automated planning for anatomical fetal brain MRI on 0.55T}

\author[1,2]{Sara Neves Silva}
\author[1,3]{Sarah McElroy}
\author[1,2]{Jordina Aviles Verdera}
\author[1,2]{Kathleen Colford}
\author[1,2]{Kamilah St Clair}
\author[1,3]{Raphael Tomi-Tricot}
\author[1,2]{Alena Uus}
\author[4]{Valéry Ozenne}
\author[2,5]{Megan Hall}
\author[2,5]{Lisa Story}
\author[2]{Kuberan Pushparajah}
\author[1,2]{Mary A Rutherford}
\author[1,2]{Joseph V Hajnal}
\author[1,2,6]{Jana Hutter}

\affil[1]{Centre for the Developing Brain, School of Biomedical Engineering \& Imaging Sciences, King’s College London, London, UK}
\affil[2]{Biomedical Engineering Department, School of Biomedical Engineering \& Imaging Sciences, King’s College London, London, UK}
\affil[3]{MR Research Collaborations, Siemens Healthcare Limited, Camberley, United Kingdom}
\affil[4]{CNRS, CRMSB, UMR 5536, IHU Liryc, Université de Bordeaux, Bordeaux, France}
\affil[5]{Department of Women \& Children's Health, King's College London, London, UK}
\affil[6]{Smart Imaging Lab, Radiological Institute, University Hospital Erlangen, Erlangen, Germany}

\corraddress{Sara Neves Silva, King's College London, London, UK}
\corremail{sara.neves\char`_silva@kcl.ac.uk}

\runningauthor{Sara Neves Silva et al.}

\begin{document}

\maketitle

\begin{abstract}

\noindent \textbf{Purpose:} Widening the availability of fetal MRI with fully automatic real-time planning of radiological brain planes on 0.55T MRI.\\
\noindent \textbf{Methods:} Deep learning-based detection of key brain landmarks on a whole-uterus EPI scan enables the subsequent fully automatic planning of the radiological single-shot Turbo Spin Echo acquisitions. The landmark detection pipeline was trained on over 120 datasets from varying field strength, echo times and resolutions and quantitatively evaluated. The entire automatic planning solution was tested prospectively in nine fetal subjects between 20 and 37 weeks. Comprehensive evaluation of all steps, the distance between manual and automatic landmarks, the planning quality and the resulting image quality was conducted.
\textbf{Results:} Prospective automatic planning was performed in real-time without latency in all subjects. The landmark detection accuracy was 4.21$\pm$2.56 mm for the fetal eyes and 6.47$\pm$3.23 for the cerebellum, planning quality was 2.44/3 (compared to 2.56/3 for manual planning) and  diagnostic image quality was 2.14 compared to 2.07 for manual planning.\\
\noindent \textbf{Conclusions:} Real-time automatic planning of all three key fetal brain planes was successfully achieved and will pave the way towards simplifying the acquisition of fetal MRI thereby widening the availability of this modality in non-specialist centres.

\end{abstract}

\section*{Introduction}
MRI plays an increasing role in both clinical antenatal diagnosis and research, complementing ultrasound (US) screening for a range of suspected fetal pathologies. The most common clinical indication \cite{Alford2016} for fetal MR imaging are thereby suspected brain anomalies. MRI has demonstrated specific additional utility when compared with US in evaluating the posterior fossa \cite{Prayer2011}, midline structures, the cortex in the progressively ossifying fetal skull, evidence of haemorrhage \cite{Epstein2222}, cysts, cleft palates \cite{Van_der_Hoek-Snieders2020-jc} and head and neck tumours \cite{Cornejo2020}, among others.

MRI offers higher spatial resolution, enhanced soft tissue contrast, and a wide range of functional contrasts. However, it poses unique challenges, such as safety considerations, involuntary fetal motion, artifacts arising from air-tissue boundaries and variability in fetal position and maternal surroundings. The most widely used acquisition technique is T2-weighted 2D single-shot Turbo Spin Echo (ssTSE), providing excellent contrast and an in-plane resolution of 1-1.5 mm, while effectively freezing motion for each individual slice.

\noindent Radiological assessment requires high-quality images in a set of defined fetal brain planes to perform a set of key measurements including the bi-parietal diameter, trans-cerebellar diameter, and ventricle diameters among others, and to visualize essential structures like the corpus callosum \cite{kyriakopoulou_normative_2017} and cerebellar vermis to identify deviations from normal development. Furthermore, gyrification and cortical development may be visualised in detail. Given the size of the studied structures, the accuracy of these measurements is crucial and relies on the availability of exact sagittal, coronal, and axial slices. While recent advances in Slice-to-Volume-Reconstruction (SVR) \cite{Gholipour2017, Uus2020, Kuklisova2012, Uus2023Bounti} allow 3D reconstructed high-resolution 3D volumes and thus re-orientation to true brain anatomy in cases where sub-optimal, oblique native planes are acquired, this technique is currently only available in specialist centres and often performed offline. Specialized radiographers trained to perform this challenging planning for fetal MRI are essential. Typically, acquisition planning is performed by optimizing the angles manually and iteratively on stacks acquired in different orientations such as whole uterus sagittal and coronal ssTSE sequences. 

\noindent Another recent development is the rediscovery of low-field 0.55T MRI. It addresses some of the aforementioned challenges of fetal MRI: the increased field homogeneity reduces distortion artifacts, the larger bore size widens access to pregnant women with larger body mass index and in later gestation, the reduced heating allows more efficient acquisitions and the longer T2* is beneficial for widely performed T2* relaxometry measurements. Recent studies showed its benefits as a promising tool for fetal MRI \cite{Aviles2023, Ponrartana2023, Payette2023}. The reduced cost, footprint, and eliminated need for shimming tools particularly carry the potential to widen access to this modality. However, a careful balance between these benefits and maintenance of adequate image quality given the reduced signal-to-noise ratio (SNR) is required and often leads to a choice of thicker slices to increase the SNR. Thicker slices, however, put further emphasis on accurate planning.

\noindent Recent work highlighted the use of AI methods during the acquisition in motion detection, correction and automatic planning. Specifically in fetal MRI, work showed the ability to perform quality control \cite{Gagoski2022}, automatic segmentation \cite{Faghihpirayesh2022, Salehi2018}, and automatic tracking \cite{Singh2020, Neves_Silva2023}. Automatic field-of-view prescription was shown in the abdomen using deep learning segmentations \cite{Lei2023-wf} and in the heart using tracking based on landmarks \cite{Xue2021-co} with successful detection ratings of 99.7\%-100\% for cine images and Euclidean distances between manual and automatically detected labels from 2 to 3.5 mm. Similar work in the brain detecting landmarks such as the anterior and posterior commissures and subsequently the symmetry line using multi-task deep neural networks were demonstrated among others by Yang et al \cite{Yang2020-nm}. Specifically for fetal MRI, segmentation of the eye region and detection of the general head position was suggested by Hoffmann et al \cite{Hoffmann2021-hd} using classical image processing methods such as maximally stable extremal regions and by Xu et al \cite{Xu2019-ls} using convolutional networks - both applicable for real-time slice planning in the future.

\noindent Here, we present an automatic, fast landmark detection and subsequent automatic radiological planning of fetal brain scans. The technique was implemented, tested, and evaluated on prospective 0.55T low-field fetal MRI scans.

\section*{Methods}
\noindent A method allowing automatic planning of the three radiological fetal brain planes using landmarks is presented and compared to manual planning. The methods section will first detail the conventional manual planning and then describe the proposed new method.

\subsubsection*{Manual planning of brain acquisitions}

\begin{figure}[ht]
     \centering
     \includegraphics[width=0.8 \textwidth]{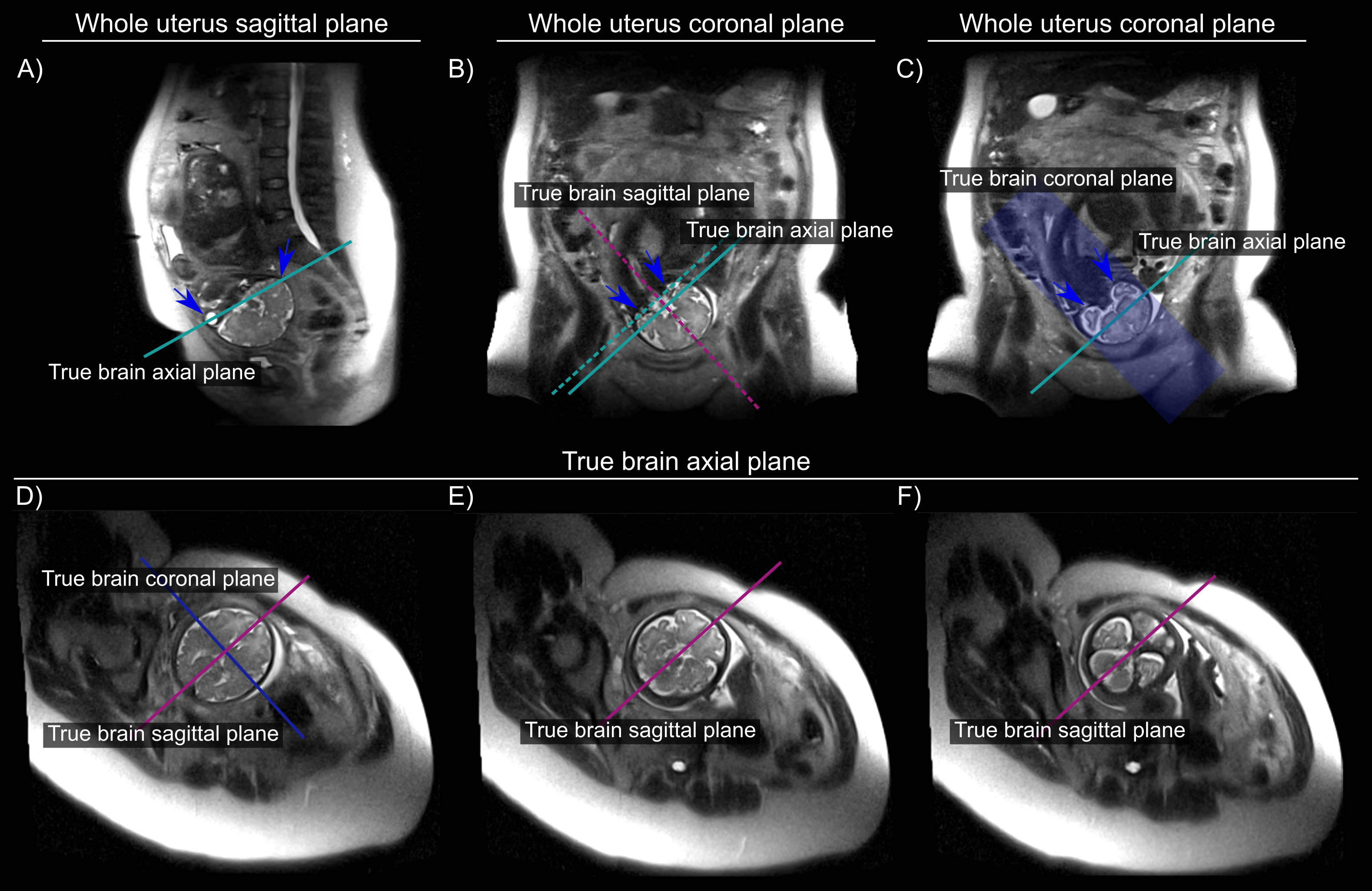}
 \caption{Conventional manual planning of the three radiological orientations for fetal anatomical T2-weighted single-shot Turbo Spin Echo (ssTSE) acquisitions. A-C illustrate whole uterus ssTSE slices in sagittal and coronal orientation to the uterus overlaid with the planning lines (cyan) to obtain (D) an axial brain ssTSE stack, allowing together with the uterus stacks to plan true sagittal (pink planning lines) and coronal (dark blue planning lines) brain ssTSE stacks. The blue arrows in A-C illustrate the used landmarks during manual planning.} \label{radiographer_planning}
 \end{figure}

\noindent Planning radiological fetal brain planes typically requires whole uterus ssTSE acquisitions as a starting point. An example is given in Figure \ref{radiographer_planning}. The orbits of the eyes and the back of the skull (Figure \ref{radiographer_planning}A, blue arrows) are used to define the orientation of the true brain axial scan, with final adjustments accomplished using the bottom of the lobes (Figure \ref{radiographer_planning}B, blue arrows, dotted blue line), slightly shifting the centre of the acquisition towards the centre of the brain (Figure \ref{radiographer_planning}B, blue line showing through-plane view). The midline between the two hemispheres allows to guide the true brain sagittal orientation (Figure \ref{radiographer_planning}B, dotted pink line, through-plane view). A favourable view of the lobes is shown in Figure \ref{radiographer_planning}C, which is subsequently used to define the true brain coronal acquisition (blue box, in-plane view). After the true brain axial scan is acquired, final adjustments to the coronal and sagittal radiological planes are performed, for instance by checking that the sagittal centre of acquisition matches the brain midline across multiple axial slices (Figure \ref{radiographer_planning}D-F, pink lines showing sagittal through-plane view). The dotted lines show preliminary planning and the full lines show the final planning of the scan.

\subsubsection*{Automatic planning of brain acquisitions}

\begin{figure}[ht]
     \centering
     \includegraphics[width=0.85 \textwidth]{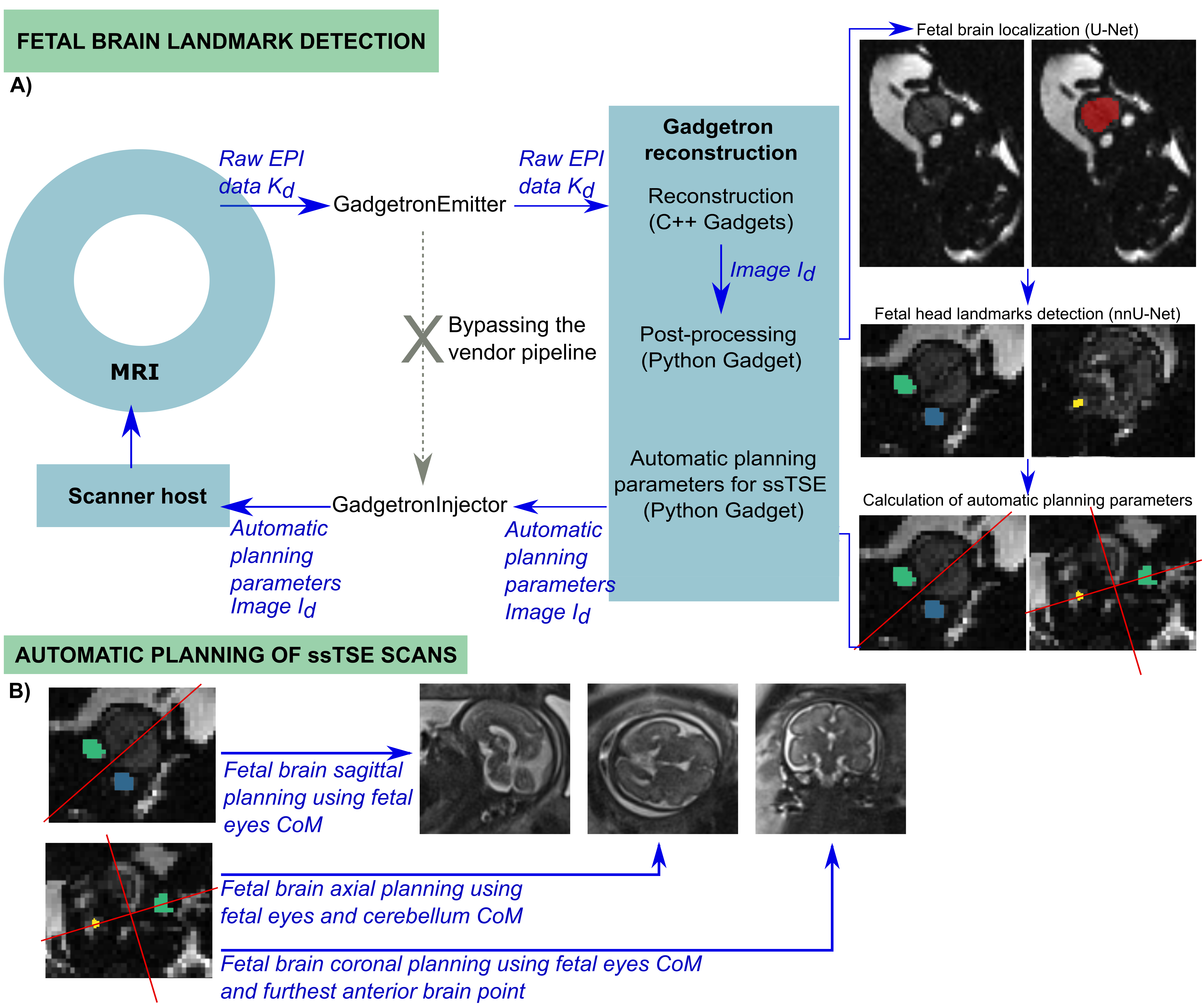}
 \caption{Schematic overview over the entire pipeline for automatic planning. (A) Detection of the fetal brain landmarks in the EPI sequence and (B) Planning parameters applied in the ssTSE sequence.} \label{fig_overview}
 \end{figure}

The automatic sequence planning framework was implemented on a 0.55T scanner (MAGNETOM Free.Max, Siemens Healthcare, Erlangen, Germany). A whole-uterus multi-echo gradient-echo single-shot echo planar imaging (EPI) sequence in coronal maternal orientation is acquired in less then 30 seconds to perform T2* mapping in several fetal organs and the placenta. In this study, this sequence is used in addition for the automatic detection of fetal brain structures enabling the automatic planning. The proposed solution consists of three steps: (A) The fetal brain position is detected on the EPI scan and a bounding box is generated, (B) The landmarks in the fetal head are identified, (C) The true fetal brain planes are calculated and applied to the following high-resolution ssTSE sequences (see Figure \ref{fig_overview} for an overview of the process and Table \ref{table_parameters} for all used imaging parameters). 

\begin{table}[h!]
\centering
\begin{tabular}{|l|l|c|c|} 
\rowcolor{gray!40}\textbf{Dataset}&\textbf{Parameters}&\textbf{Subjects}&\textbf{Time}\\
\rowcolor{white}\textbf{EPI training} 1.5T& 1.5T Philips Ingenia, 28-channel torso coil&&\\
\rowcolor{white}& Matrix=144x144-288x288, Resolution 2.5 mm isotropic, 30–96 slices&&\\
\rowcolor{white}&TE=[14.6/77.4/140.1/202.8/265.5] ms & 80 & 22.5 s \\ 
\hline
\rowcolor{white}\textbf{EPI training} 3T& 3T Philips Achieva, 32-channel cardiac coil&&\\
\rowcolor{white}& Matrix=144x144-192x192, Resolution 2/3 mm isotropic, 45–75 slices &&\\
\rowcolor{white}&TE=[3.8/70.4/127/183.6] / [10.1/54.3/98.4/142.5/186.8] ms& 77 & 18.3 s \\ 
\hline
\rowcolor{gray!20}\textbf{EPI autoplan}& 0.55T Siemens MAGNETOM Free.Max, 6-channel coil, 9-channel spine coil &&\\
\rowcolor{gray!20}&Matrix=100x100-128x128, Resolution 3.13–4.0 mm isotropic, &&\\
\rowcolor{gray!20}&TE=81 ms, 50–59 slices& 9 & 25 s \\ 
\hline
\rowcolor{gray!20}\textbf{ssTSE autoplan}& 0.55T Siemens MAGNETOM Free.Max, 6-channel coil, 9-channel spine coil&&\\
\rowcolor{gray!20}&Matrix=304x304, Resolution 1.48x1.48x4.5 mm\textsuperscript{3}, 35 slices&&\\
\rowcolor{gray!20}& TE=106 ms, TR=14.6 s & 9 & 55 s \\ 
\end{tabular}

\caption{Sequence parameters for all described sequences used for training (first two rows, white) and for the prospective automatic planning method (last two rows, gray).}
\label{table_parameters}
\end{table}

\subsection*{(A) Localization of the fetal brain}
First, the EPI sequence was modified to export the acquired raw data to a Gadgetron reconstruction pipeline \cite{Hansen2013}, deployed on an external GPU-equipped (NVIDIA GEFORCE RTX 2080 Ti, NVIDIA Corporate, Santa Clara, CA) computer connected to the internal network of the MRI scanner. The raw data is converted to ISMRMRD format immediately upon acquisition and reconstructed using off-the-shelf Gadgets that provide generic modules for configuring the streaming reconstruction in the Gadgetron framework. Then, a Python Gadget was implemented to automatically estimate the position of the region of interest in the image using a pre-trained 3D UNet \cite{Uus2021UNET} for fetal brain localization. A bounding box encompassing the fetal brain is calculated and used for the following landmark detection task \cite{Neves_Silva2023}. 

\paragraph{Datasets and training}
The fetal brain localization network was trained on 125 labelled fetal EPI datasets acquired at 1.5T/3T and tested on 29 0.55T fetal datasets. To increase the robustness of the network, the EPI scans used for the training and performance evaluation of the model deliberately vary in the acquisition parameters (field strength, echo time, resolution, acceleration factor), gestational age (15-40 weeks), fetal health (control cases, fetal growth restriction, prolonged preterm rupture of the membranes, etc.), and fetal position (cephalic, breech, transverse). The parameters for the training data set were: 1) 3T Philips Achieva, 32-channel cardiac coil and 16-channel spine coil, matrix size=144x144-192x192, isotropic resolution 2mm\textsuperscript{3}, TE=[3.8/70.4/127/183.6]/ [10.1/54.3/98.4/142.5/186.8] ms, 45–75 slices; 2) 1.5T Philips Ingenia, 28-channel torso coil, matrix size=144x144-288x288, isotropic resolution2.5mm, TE=[14.6/77.4/ 140.1/202.8/265.5] ms, 30–96 slices. 
The trained model was tested on low-field fetal datasets acquired on a 0.55T Siemens MAGNETOM Free.Max, using a blanket-like BioMatrix Contour-L 6-channel coil and fixed 9-channel spine coil, matrix size=100x100-128x128, isotropic resolution 3.13–4.0 mm, TE=[46/120/194/268/342] ms, 50–59 slices. 

\subsection*{(B) Landmark detection}
\begin{figure}[ht]
     \centering
     \includegraphics[width=0.6 \textwidth]{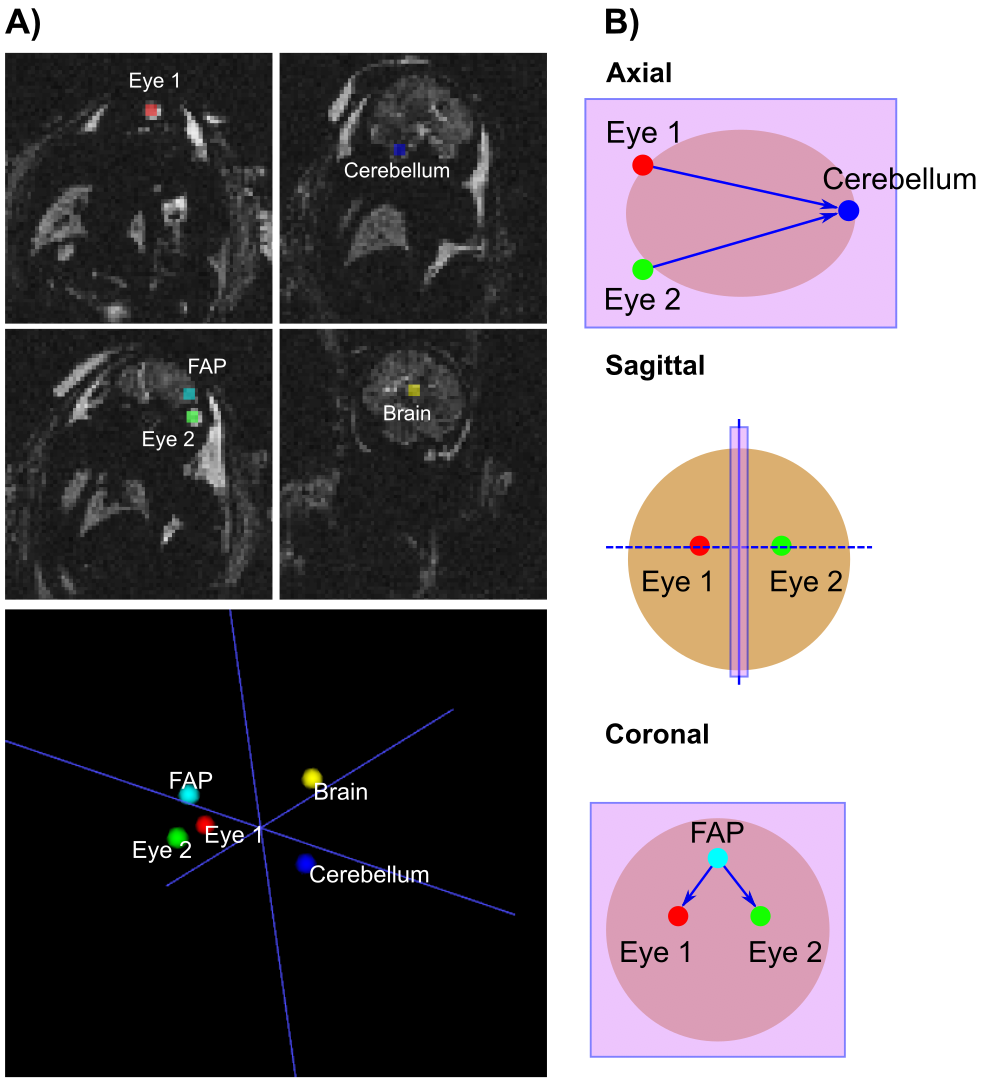}
 \caption{A) Key points extracted from the whole-uterus coronal EPI scan for automatic planning of the ssTSE scans are illustrated: these include key landmarks eyes (red and green dots) and cerebellum (blue dot) and two points extracted from the fetal brain: the brain centre of mass (CoM, yellow dot) and the furthest anterior point. B) The axial orientation is defined using the eyes and cerebellum centre of mass coordinates, the sagittal orientation uses the eyes CoM coordinates, and the coronal orientation is defined using the eyes CoM and furthest anterior point of the fetal brain mask. The brain CoM is used to define the centre of the slice stack for all radiological planes. The purple boxes illustrate the slice position for each orientation: in axial and coronal planes the views shown are in-plane, and for the sagittal plane the view shown is through-plane.} \label{fig_geometry}
 \end{figure}

A deep learning-based landmark detection method using the nnUNet framework \cite{Isensee2021} was adopted to extract specific head landmarks, concretely the orbit of both fetal eyes and the lower edge of the cerebellum. This framework performs semantic segmentation and it automatically adapts to a given dataset by analysing the provided training data and configuring a matching UNet-based segmentation pipeline. Furthermore, the furthest anterior point (FAP) of the fetal brain mask is extracted. The centre of mass (CoM) coordinates of the landmarks, the brain mask, and the FAP are then transformed into the patient coordinate system and written into a text file on the external server and scanner host. Figure \ref{fig_geometry}A illustrates the key points extracted and used for planning the radiological acquisitions.

\paragraph{Datasets and training}
The landmark nnUNet was trained and tested on cropped and labelled images of 76/15 fetal subjects, respectively. Automatic cropping was performed by first applying the inference of the fetal brain localization network to the datasets and using the estimated fetal brain position to define a bounding box that is defined by the brain mask and an expansion factor of 50\%, defining how much the bounding box should be enlarged relative to the dimensions of the fetal brain, ensuring that the cropped region contains all key landmarks. The gold standard segmentations of the landmarks were manually drawn for each dataset. All slice stacks were acquired using the low-field fetal protocol described above.

\subsection*{(C) Calculation of the radiological brain planes}

\subsubsection*{Automatic orientation calculation of brain acquisitions}
\noindent The ssTSE sequence that follows the EPI sequence in the protocol was modified to use the information stored in the file to calculate the three standard radiological planes axial, sagittal, and coronal. The normal vector to the sagittal plane is calculated as the vector between the eyes landmarks (see Figure \ref{fig_geometry}B). It was identified that the eyes and FAP lie on the same approximately coronal plane and the eyes and cerebellum lie on the same approximately axial plane. Therefore, the normal vector to the coronal plane was calculated as the dot product of the two vectors between each eye and the FAP, and the normal vector to the axial plane as the dot product of two vectors between each eye and the cerebellum. A hierarchical approach is chosen to increase robustness even more: If the cerebellum is not extracted successfully, the CoM of the brain is used instead for the vector calculation. The CoM of the brain is used to define the centre of the slice stack for all planes.

\subsection*{Experiments and evaluation}

\subsubsection*{Evaluation of the fetal brain localization and landmark detection accuracy}
\noindent The performance of the localization network was analyzed using the Dice similarity coefficient (DSC) and the Intersection-over-Union (IoU) metric calculated between the gold standard manual segmentations performed by two fetal experts (9 years and 2 years of fetal MRI experience respectively) and the brain masks generated by the network. The landmark detection performance was analyzed by calculating the 3D distance between the centre of mass of the manual segmentations of the landmarks performed by fetal MRI experts and the landmarks obtained from the network.

\subsubsection*{Real-time fetal brain plane planning}
\noindent The entire pipeline was acquired prospectively in nine pregnant volunteers in St Thomas’ Hospital, recruited between October-December 2023 after informed consent was obtained as part of two ethically approved studies (MEERKAT REC19/LO/0852 and miBirth 23/LO/0685). Women were scanned on the above-described clinical 0.55T MAGNETOM Free.Max scanner in the supine position with leg support to ensure comfort with life monitoring throughout the scan. Gestational ages ranged between 20.3 and 37.3 weeks (mean 32.8+-3.5). For each fetal subject, the initial whole-uterus coronal multi-echo gradient-echo single-shot EPI scan was acquired (resolution=3.13-4.0 mm isotropic, TE=[81/227/372] ms, 50–59 slices, matrix size=128x128). Subsequently, the automatically-planned ssTSE (autoplan-ssTSE) prototype sequence was acquired in all three radiological planes with the following parameters: resolution=1.48x1.48x4.5 mm\textsuperscript{3}, TE=106 ms, 35 slices, matrix size=304x304, TE=106 ms, TR=14.6 s. In addition, the same sequence was acquired with manual planning of the three radiological planes by fetal radiographers (1-5 years of experience) for all subjects. In total, for each fetal subject, three planes (axial, coronal, sagittal) were acquired, resulting in a total of 27 acquisitions of each planning approach - manual and automatic. The time required for planning by the radiographers and the automatic method was measured. 

\subsubsection*{Evaluation of the planning quality and diagnostic quality of the prospectively acquired images}
\noindent Quantitative evaluation was performed blinded to the method and the ability to produce accurate measures was scored. The resulting automatically planned radiological planes were assessed by a fetal radiographer with five years of fetal MRI experience answering the questions "Rate the planning quality [1-3]" (1 - re-acquisition required, 2 - usable, 3 - full brain coverage, symmetry of the brain structures, no re-acquisition required). A fetal radiologist with $>$ 15 years of experience additionally quantified the clinical value of the automatic planning method by attempting to perform, for both manually and automatically planned acquisitions, five measurements in the sagittal view (corpus callosum, cerebellar vermis, pons anterior-posterior (AP) diameter, pituitary stalk, hard palate), four in the coronal view (trans-cranial diameter (TCD), bi-parietal diameter (BPD), cavum septum pellucidum (CSP), ventricle) and five in the axial view (BPD, occipito-frontal diameter (OFD), ventricle, TCD, CSP). The radiologist was asked to answer "Can you perform all radiological measurements? [0-5]" (0 - unusable, 5 - all five measurements successfully performed). The results were in addition assessed against gestational age and maternal Body Mass Index.

\section*{Results}

\subsection*{(A) Fetal brain localization}
The brain localization task, trained on mid/high-field data, achieved an overall DSC of 0.82±0.18 and IoU of 0.73±0.19 when tested on low-field scans across all TEs, fetal positions, and gestational ages. Extraction of fetal brain masks took between 11.43 and 20.85 ms per volume in the offline testing mode.

\subsection*{(B) Landmark detection}
\begin{figure}[ht]
     \centering
     \includegraphics[width=0.75 \textwidth]{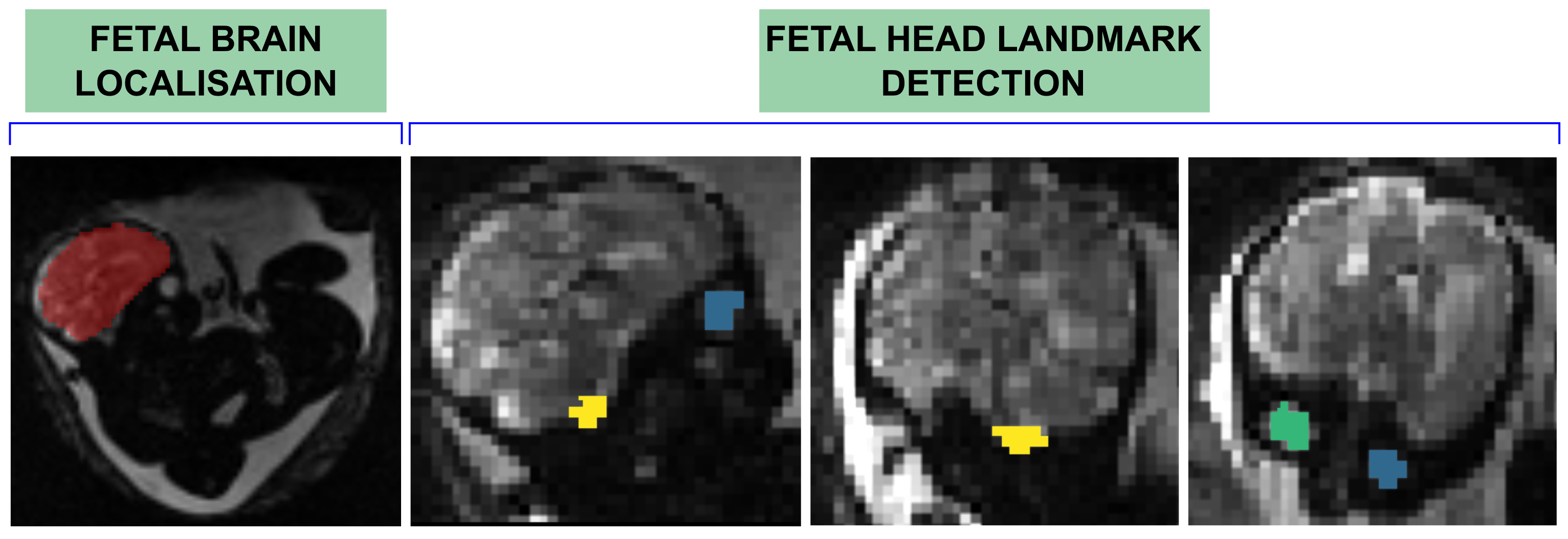}
 \caption{Two-step localisation task - global localisation of the fetal head using a fetal brain segmentation 3D UNet model, and local localisation with extraction of the 3 fetal head landmarks - eyes and cerebellum.} 
 \label{fig_example}
\end{figure}

\begin{figure}[ht]
     \centering
     \includegraphics[width=0.75 \textwidth]{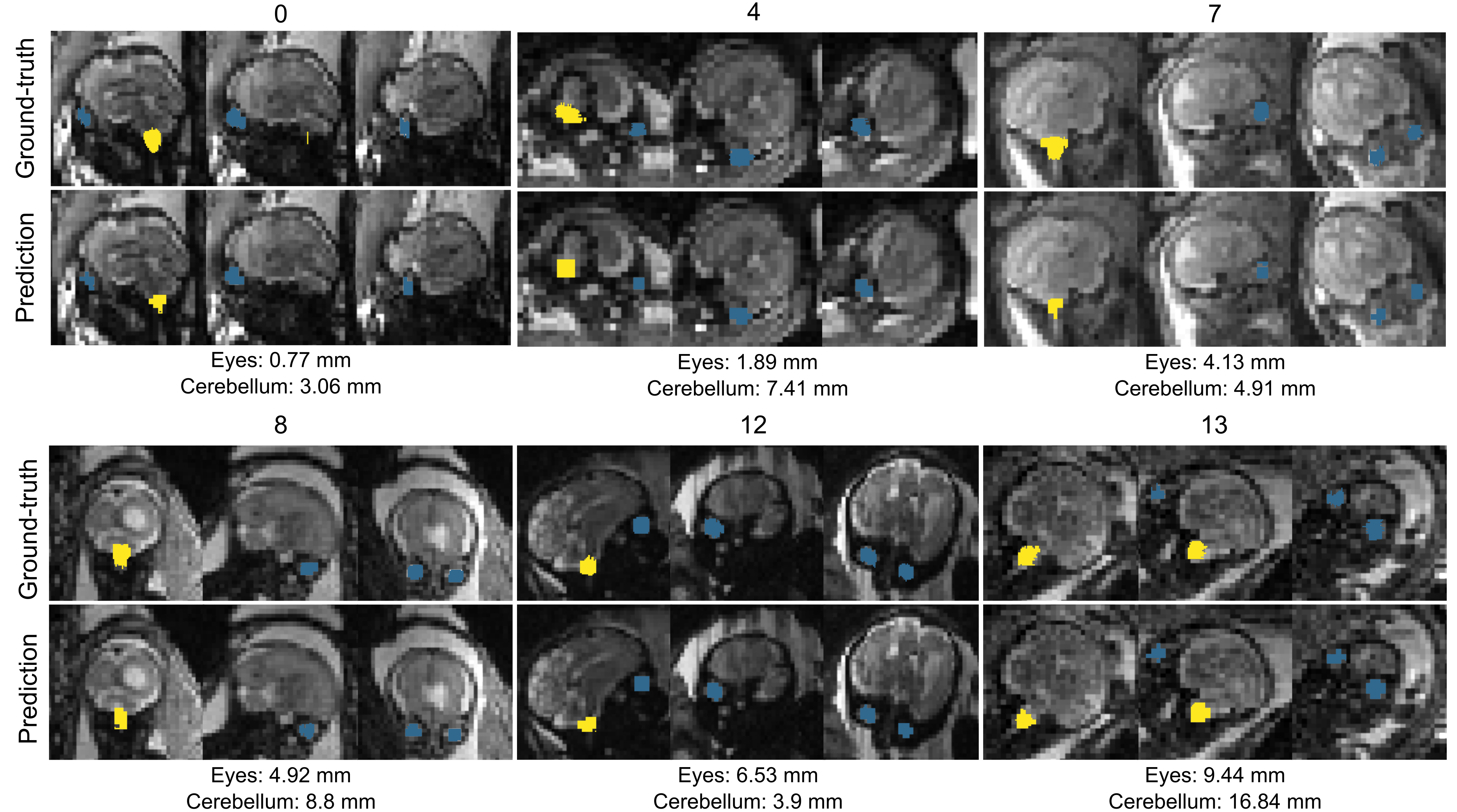}
 \caption{Ground-truth and predicted landmarks, generated by the nnUNet, for fetal subjects 0, 4, 7, 8, 12, and 13 of the landmark detection test set. The distance between the centre of mass of each is additionally displayed. The model outputs the eyes landmarks using a single label, hence they are depicted with the same colour (and label). This result is then processed to separate the two unconnected regions to be used to guide the planning of the ssTSE scan.} 
 \label{fig_landmarks_results}
\end{figure}

The mean distance between the CoM of the eyes and cerebellum labels from automatic and manual segmentations was 4.21$\pm$2.56 mm and 6.47$\pm$3.23 mm, respectively. Figure \ref{fig_example} shows the predicted brain and landmarks segmentations for one fetal subject of the landmarks model test set. Figure \ref{fig_landmarks_results} shows examples of predicted landmarks generated by the nnUNet, the corresponding ground-truth segmentations, and the distance between the two. Landmark detection took 5.18 seconds per volume in the offline testing mode.

\subsection*{(C) Real-time fetal brain plane planning}
\begin{figure}[ht]
     \centering
     \includegraphics[width=.75 \textwidth]{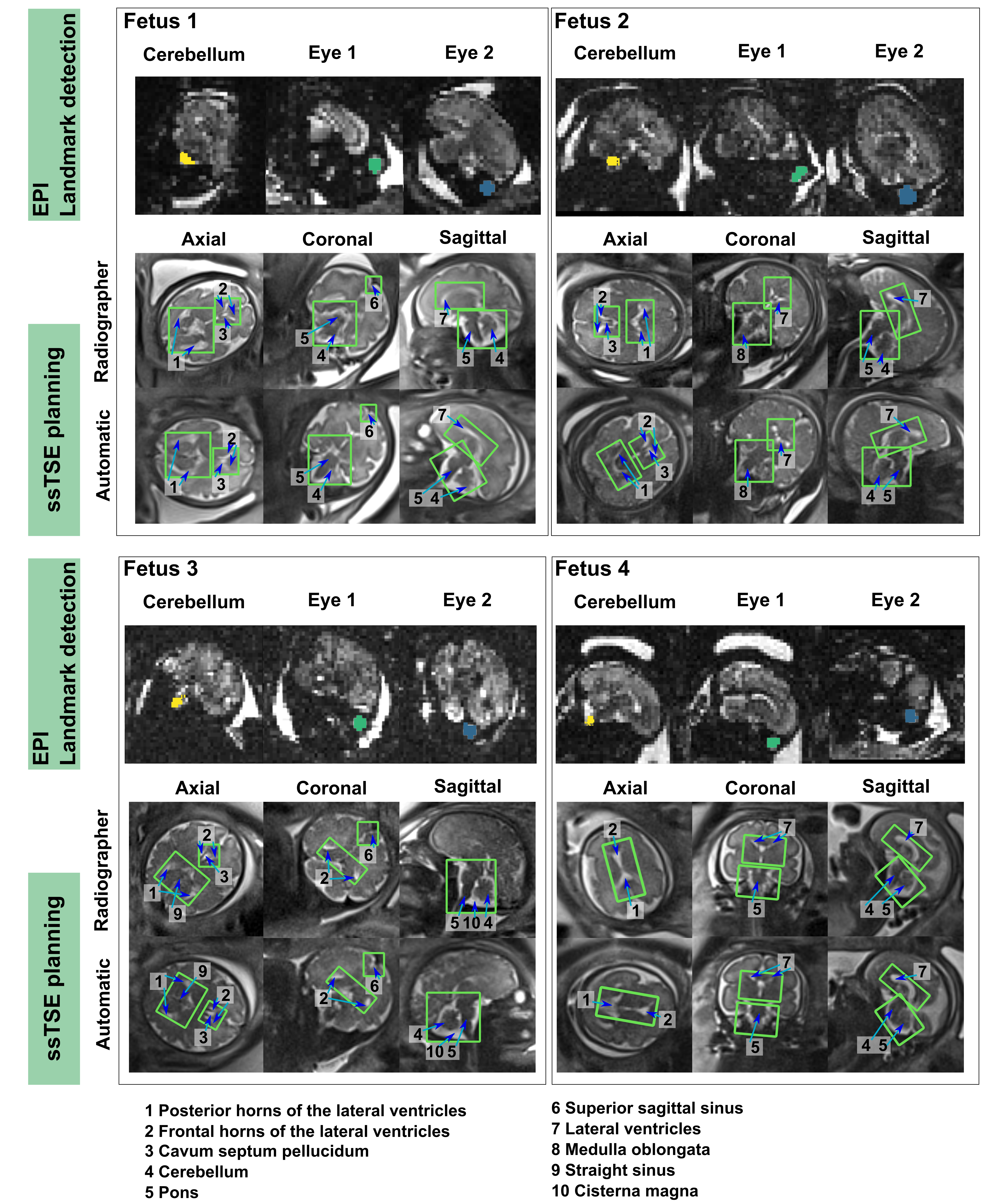}
 \caption{Landmarks extracted from whole uterus coronal EPI scans and radiographer-planned and automatic (orientation defined from the position of the eyes and cerebellum, extracted from the EPI) radiological ssTSE acquisitions. Anatomical structures of interest were highlighted (green squares, blue arrows) for manual and automatic acquisitions to demonstrate comparable orientation planning.}
 \label{fig_prospective}
\end{figure}

The entire pipeline was successfully run in all nine fetal scans, with all steps of the approach successfully implemented and tested prospectively. The estimated time between the completion of the EPI sequence and the start of the ssTSE sequence, ready to be run with the planning parameters correctly set, is less than 5 seconds with the automatic method, and an average of 1.5 minutes for the specialist radiographer and 4 minutes for the novice radiographer. Landmark detection and automatic radiological planning are shown in Figure \ref{fig_prospective} alongside the radiographer's manually planned acquisitions, with matching anatomical structures in the manual and automatic acquisitions highlighted to demonstrate that both techniques allow the same structures of interest to be captured comparably. The anatomies (green squares, blue arrows) are labelled between 1-10, with the corresponding dictionary at the bottom.

\begin{figure}[ht]
     \centering
     \includegraphics[width=1. \textwidth]{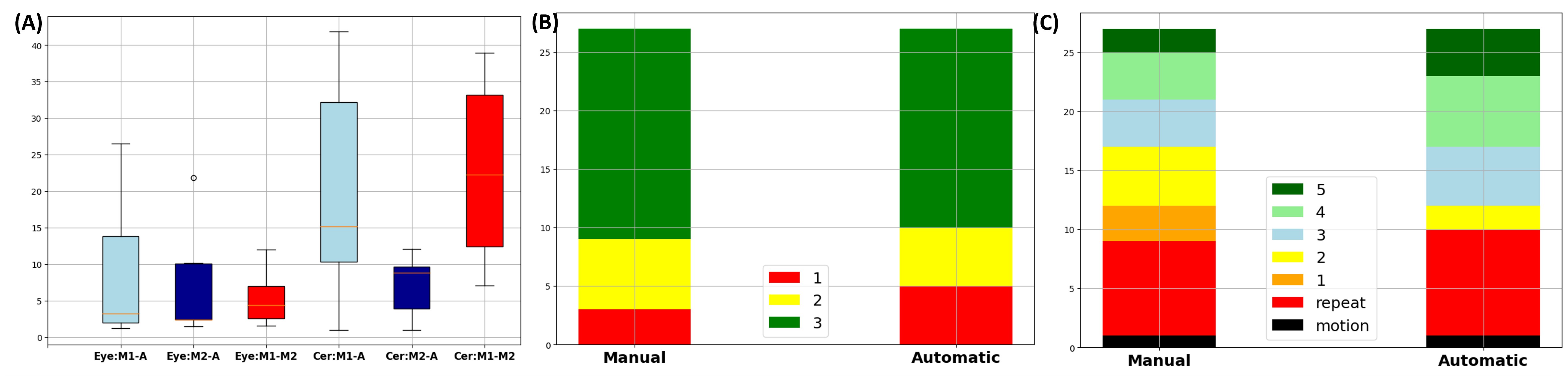}
 \caption{Quantitative evaluation of the 27 prospectively acquired ssTSE data sets. (A) Landmark assessment: Difference in mm between the manual segmentations performed by two observers (M1 and M2) and the automatic method (A) for the eyes and the cerebellum. (B) Radiographer assessment: Image quality for all 27 acquisitions for the manual and automatic acquisitions. (C) Radiologist assessment: Image quality for all 27 acquisitions for the manual and automatic acquisitions. }
 \label{fig_boxpl}
\end{figure}

In Figure \ref{fig_boxpl}A, a quantitative assessment of the real-time landmark detection task applied to the prospective whole uterus EPI scans is depicted. The landmarks were extracted in real-time during the scan to automatically plan the radiological brain acquisitions and were subsequently segmented manually by two observers M1 and M2, where M2 is the observer that segmented the landmarks for training the nnUNet. All landmarks but one cerebellum segmentation (fetal subject 9 of GA 20 weeks) were prospectively extracted during the scan. The inability of the model to extract the landmark may be linked to the early gestation of the fetus and the consequent small size of the brain structures, as well as the fact that early gestation fetuses are not as well represented in the training dataset of the model, a natural bias that is often present in fetal MRI studies. For this case, the CoM of the brain was used instead as a landmark to allow for the automatic planning to still be performed. For the 9 fetal subjects scanned, the differences in distances (in mm) between the CoM of eyes and cerebellum landmarks segmented by M1, M2, and automated (A) were calculated. For the eyes, the difference between M1 and A ranged between 1.25-21.81 mm, M2-A between 1.47-10.13 mm, and M1-M2 between 1.54-11.98 mm. As for the cerebellum, the CoM differences in distance for M1-A ranged between 9.43-41.88 mm, M2-A between 1.99-12.1 mm, and M1-M2 between 7.11-38.91 mm. While inter-observer variability never exceeded 11.98 mm for the eyes landmarks, the cerebellum presented much larger variability in the segmentations produced by M1 and M2, which can be explained by the boundaries of this anatomical structure not being as well defined as for the eyes, leading to variability in the size of the segmentation. For this reason, using a deep learning approach helps to eliminate observer bias, as segmentations are more standardised and consistent.

\begin{figure}[ht]
     \centering
     \includegraphics[width=1. \textwidth]{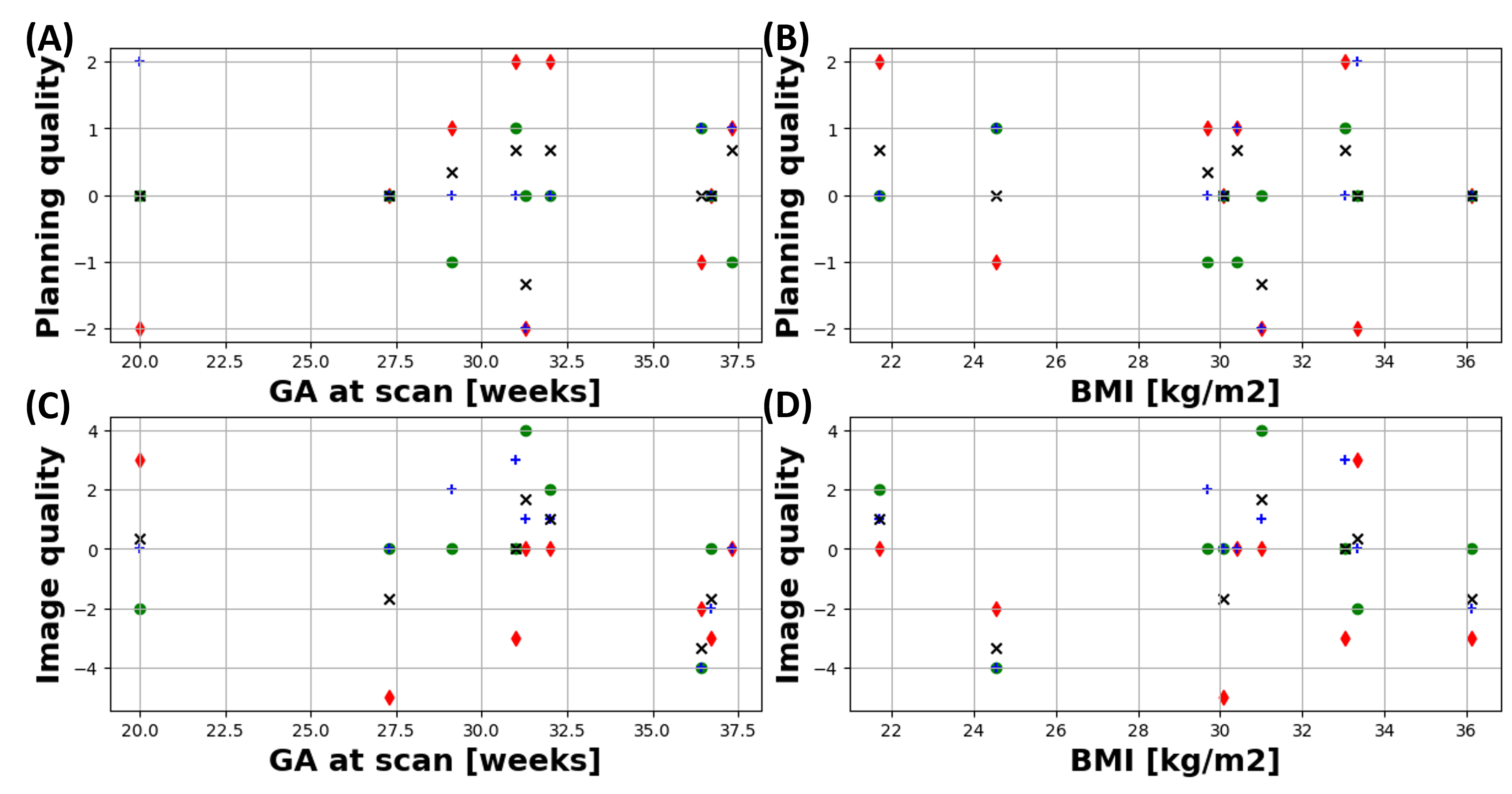}
 \caption{Quantitative assessment of the planning quality (A-B) and image quality (C-D) illustrated for all three planes (coronal, axial, and sagittal) against GA (A, C) and maternal BMI (B, D).}
 \label{fig_gabmi}
\end{figure}

The quantitative evaluation of the three achieved radiological planes is displayed in Figure \ref{fig_boxpl}B and C and Figure \ref{fig_gabmi}. In both Figures, the assessments performed by the radiographer and radiologist aimed to score how close to a perfect true brain plane each acquisition was and the clinical value of the automatically planned images (compared with the manually planned ones). With 3 radiological planes (axial, coronal, sagittal) acquired for each fetal subject, 27 planes were acquired in total in this work. Out of 27 automatically planned acquisitions, 22 scans were successfully acquired without the need for re-acquisition, and 24 out of 27 manually planned scans similarly did not require repeating according to the analysis performed by a fetal radiographer. Full brain coverage was achieved in all radiological planes for all manual and automatic scans, thus the need for re-acquisition resided in large tilts present in the orientation of the brain, with asymmetries visible between the two brain hemispheres and anatomical structures of interest present in both. 

In 27 manually planned scans, 6 scans displayed slight asymmetry with small rotations of the anatomy that would not affect the clinical value of the images and thus did not require re-acquisition (4 coronal planes and 2 axial). Severe asymmetry was observed in 3 acquisitions (2 axial planes and 1 sagittal) with re-acquisition needed. The remaining 18 acquisitions showed a complete symmetry of the two brain hemispheres. Regarding the automatic planning method, full symmetry of the brain hemispheres and neurological structures was achieved in 17 automatic acquisitions, demonstrating the ability of the method to plan true brain planes accurately. Generating the automatic acquisition plane took, on average, 5 seconds, and the radiological acquisitions were acquired in less than one minute each. Asymmetry was observed in 10 cases, although at different degrees: 5 scans showed larger asymmetry with severe tilts (3 axial planes, 1 coronal, and 1 sagittal), and 5 showed smaller tilts causing slight asymmetry which would not affect the clinical value of acquisitions and therefore would not require re-acquisition (2 coronal planes, 2 sagittal, and 1 axial). Figure \ref{fig_boxpl}B illustrates the radiographer assessments. In summary, a quality score of 81.5\% (2.44/3) was achieved by the automatic planning method and 85.2\% (2.56/3) for manual planning.

The clinical value and image quality were additionally evaluated by a fetal radiologist. With a maximum score of 14 measurements per fetal subject, the number of measurements successfully performed in the manually planned acquisitions was: 14, 10, 7, 6 (2 subjects), 5 (2 subjects, 1 of GA 20 weeks), 2, and 1. Artefacts were also visible in 3 coronal acquisitions, which may have affected the ability to perform measurements. Using the automatic planning method, the number of measurements performed was: 12, 11 (2 subjects), 7 (one plane showing artefacts), 5 (2 subjects), 4 (fetal subject of GA 20 weeks), 3, and 0 (one plane showing distortion). An average of 2.07/5 measurements were successfully performed by a radiologist for the manual acquisitions, and 2.14/5 for the automatic acquisitions. 

The measurements with the highest scores across all fetal subjects and all orientations were the CSP and the ventricle, in the coronal orientation, with scores of manual 8/9 and automatic 6/9 subjects for both measurements. The corpus callosum (manual 1/9, automatic 0/9 subjects) and pons AP diameter (manual 1/9, automatic 2/9 subjects), both in sagittal orientation, showed the lowest scores. For both sagittal and coronal orientations, radiographer-planned acquisitions scored slightly higher than the automated ones (by 2.22\% and 5.0\%, respectively), however, the automatically-planned axial orientation was superior to the manual planning by 11.1\%. Overall, the automatic method for planning the acquisitions was 1.59\% higher than the manual planning approach, according to the image quality quantitative assessment performed by a radiologist.

As illustrated in Figure \ref{fig_boxpl}C, out of 27 manual scans 5 measurements were performed in 2 scans, 4 measurements in 4 scans, 3 in 4 scans, 2 in 5 scans, 1 in 3 scans, and 1 dataset severely affected by motion, resulting in 8 scans without immediate clinical value (unless tools for rotating the brain anatomy are utilised). Regarding the automatically-planned scans, 5 measurements were performed in 4 scans, 4 measurements in 6 scans, 3 in 5 scans, and 2 in 2 scans. One dataset was similarly corrupted by motion, resulting in 9 scans without immediate clinical value.

Figure \ref{fig_gabmi} depicts planning quality, assessed by a radiographer, and image quality, assessed by a radiologist, according to gestational age (GA at scan) and maternal Body Mass Index (BMI).

\section*{Discussion and Conclusion}
In this study, a fully automated framework for planning all three radiological fetal brain planes was demonstrated in low-field MRI scans. The method presented uses a whole-uterus coronal multi-echo EPI sequence, acquired and employed for T2* relaxometry, to serve the additional purpose of fetal head landmark detection. The brain and key landmarks are automatically extracted and handed to the subsequent ssTSE acquisition to plan fetal brain-oriented acquisitions in under 5 seconds. This automatic planning of fetal MRI scans, shown here as comparable to the manual planning for clinical reporting, reduces dependence on specialist staff and increases time efficiency and thus carries the potential to significantly widen accessibility to fetal MRI beyond specialist centres.

In line with recent work \cite{Hoffmann2021-hd}, the eyes were successfully used as landmarks and were crucial for calculating the orientation of the fetal brain, however, the detection shown here did not include further pose estimation or image processing steps. While recent work employing AI to detect landmarks and organs on the MR acquisitions focused on 1.5T and 3T \cite{Salehi2019, Neves_Silva2023, Hoffmann2021-hd, Xue2021-co, Xu2019-ls, Gagoski2022}, the present work was performed on low field fetal MRI at 0.55T, providing increasing challenges regarding SNR and resolution, but allowing to take key steps towards wider accessibility of fetal MRI.

The retrospective evaluation of the network that extracts the head landmarks (eyes and cerebellum), resulted in an average distance between the CoM of the manually segmented and automatically extracted eyes of 4.21 mm, which corresponds to 1.34 voxels, and 6.47 mm for the cerebellum, corresponding to 2.07 voxels, which is sufficiently robust for the calculation of the radiological planes. Fetal subjects 8 and 13 of the test set showed the largest distances (mm) between the CoM of manual and automatic landmark masks, however, Figure \ref{fig_landmarks_results} demonstrates how even at such large distances (9.44 mm and 16.84 mm for eyes and cerebellum of Fetus 13) the predictions are still highly comparable to the gold standard segmentations and extract the key landmarks accurately enough for the orientation calculations of the ssTSE scans. According to the quantitative and qualitative assessments of the calculation of the radiological brain planes, the automatically planned scans achieved comparable image quality to the scans manually planned by an experienced fetal radiographer.

\noindent Major strengths of the current study are the complete and robust real-time deployment. The two-step approach presented here for landmark detection: first the identification of the brain with a 3D UNet followed by the landmark detection using nnUNet and the hierarchical application of landmarks in cases where individual landmarks are not identified - e.g. replacing the cerebellum by the centre of the brain for the sagittal prescription - contributes to robustness even in younger fetuses.
Furthermore, the developed method was built using open-source frameworks, facilitating straightforward translation and dissemination. All steps, including the EPI reconstruction with off-the-shelf Gadgets, the fetal brain localization and landmark detection trained models, and Gadget are available to any interested researchers (https://github.com/saranevessilva/fetal-brain-landmarks). All anonymized fetal data is also available upon request.

\noindent There are, however, limitations with the current study. First, the present study is a single-center study. The training data and retrospectively assessed test data are small, as well as the number of prospective cases, which currently does not allow a more robust evaluation of the method in a larger cohort. This is often the case in fetal MRI as such a modality is not yet widely available worldwide and data sharing policies between institutions and studies are often restrictive. The framework may need to be further refined to perform robustly in wider populations and across different scanners/field strengths. Future work will hence investigate further developments to ensure the robustness of the automatic planning framework in this wider context. EPI images are reconstructed using standard Fourier reconstruction computed with off-the-shelf Gadgets (without image filtering or AI-based image enhancement provided by the scanner reconstruction pipeline) resulting in limited SNR, which may hinder the performance of the landmark detection. This will be addressed by further improving the real-time reconstruction pipeline, by including image filtering or AI-based image enhancement to achieve increased SNR in the EPI images. Fetal MRI at higher field will suffer less from this limitation. Next, this framework is based on a whole-uterus multi-echo EPI sequence, not requiring any planning and fitting to the standard patient examination as current practice for all fetal MRI scans in our institution. However, the reduced resolution of this EPI sequence (3 mm isotropic) compared to the ssTSE sequence (1.25 mm in-plane) might influence the achieved precision regarding landmark detection. The network and achieved complete automatic planning can however be extended to whole uterus ssTSE localiser scans in the future. Furthermore, while the landmarks were chosen as independently as possible from brain structures involved in common pathologies such as the ventricles and the corpus callosum - and in line with recent work \cite{Hoffmann2021-hd} - the automatic planning in fetuses with developmental abnormalities in the location of the lower edge of the cerebellum or the orbits of the eyes might involve the need for additional manual adjustment. Finally, adding a fourth landmark in the back of the skull might be helpful to further stabilize the planning of the acquisition in the axial orientation.

\noindent Future work towards an automatic fetal brain MR examination will furthermore include the real-time deployment of automatic quality control \cite{Gagoski2022}, sub-regional segmentation \cite{Uus2023Bounti} and automatic biometry \cite{She2023-vl}.

\noindent To conclude, this study shows the feasibility of rapid deep learning-based automatic planning of anatomical radiological fetal brain scans and presents an open-source framework open for further extensions and improvements. The clear next application is fetal cardiac landmark detection for automatic planning of phase contrast sequences acquired to study the flow in the major arteries.

\section*{Acknowledgements}
The authors thank all pregnant women and their families for taking part in this study, the midwives Imogen Desforges, Chidinma Iheanetu Oguejiofor, and Maggie Lee, and clinical research fellow Simi Bansal and Vanessa Kyriakopoulou for their invaluable efforts in recruiting and looking after the women in this study as well as Massimo Marenzana and Michela Cleri for their involvement in the acquisition of these datasets. This work was supported by a Wellcome Trust Collaboration in Science grant [WT201526/Z/16/Z], Heisenberg funding from the DFG [502024488], a UKRI FL fellowship, an NIHR Advanced Fellowship to LS [NIHR3016640], an EPSRC Research Council DTP grant [EP/R513064/1] and by core funding from the Wellcome/EPSRC Centre for Medical Engineering [WT203148/Z/16/Z]. The views presented in this study represent those of the authors and not of Guy's and St Thomas' NHS Foundation Trust.


\newpage
\begin{thebibliography}{29}
\providecommand{\natexlab}[1]{#1}
\providecommand{\url}[1]{\texttt{#1}}
\providecommand{\urlprefix}{}

\bibitem{Alford2016}
Alford, R.E., Bailey, A.A., Twickler, D.M. (2016).
\newblock Fetal Central Nervous System.
\newblock In: Masselli, G. (eds) MRI of Fetal and Maternal Diseases in Pregnancy.
\newblock Springer, Cham.
\newblock pp. 91--118.
\newblock \url{https://doi.org/10.1007/978-3-319-21428-3_6}.

\bibitem{Prayer2011}
Prayer, D., Brugger, P.C., Nemec, U., Milos, R.I., Mitter, C., Kasprian, G. (2010).
\newblock Cerebral Malformations.
\newblock In: Prayer, D. (eds) Fetal MRI.
\newblock Medical Radiology, pp. 287--308. 
\newblock Springer, Berlin, Heidelberg.
\newblock \url{https://doi.org/10.1007/174_2010_117}.

\bibitem{Epstein2222}
Epstein, K.N., Kline-Fath, B.M., Zhang, B., Venkatesan, C., Habli, M., Dowd, D., Nagaraj, U.D.
\newblock Prenatal Evaluation of Intracranial Hemorrhage on Fetal MRI: A Retrospective Review.
\newblock American Journal of Neuroradiology, 42(12):2222--2228, 2021.
\newblock DOI: 10.3174/ajnr.A7320.
\newblock \url{https://www.ajnr.org/content/42/12/2222}.

\bibitem{Van_der_Hoek-Snieders2020-jc}
van der Hoek-Snieders, H. E. M., van den Heuvel, A. J. M. L., van Os-Medendorp, H., Kamalski, D. M. A.
\newblock Diagnostic accuracy of fetal MRI to detect cleft palate: a meta-analysis.
\newblock Eur. J. Pediatr., 179(1):29--38, January 2020.
\newblock \url{https://link.springer.com/article/10.1007/s00431-019-03526-7}.

\bibitem{Cornejo2020}
Cornejo, P., Feygin, T., Vaughn, J., Pfeifer, C. M., Korostyshevska, A., Patel, M., et al.
\newblock Imaging of fetal brain tumors.
\newblock Pediatr. Radiol., 50(13):1959--1973, December 2020.
\newblock \url{https://link.springer.com/article/10.1007/s00247-020-04664-1}.

\bibitem{kyriakopoulou_normative_2017}
Kyriakopoulou, V., Vatansever, D., Davidson, A., Patkee, P., Elkommos, S., Chew, A., et al.
\newblock Normative biometry of the fetal brain using magnetic resonance imaging.
\newblock Brain Structure \& Function, 222(5):2295--2307, July 2017.
\newblock DOI: 10.1007/s00429-016-1345-8.
\newblock \url{https://link.springer.com/article/10.1007/s00429-016-1345-8}.

\bibitem{Gholipour2017}
Gholipour, A., Rollins, C. K., Velasco-Annis, C., Ouaalam, A., Akhondi-Asl, A., Afacan, O., et al.
\newblock A normative spatiotemporal MRI atlas of the fetal brain for automatic segmentation and analysis of early brain growth.
\newblock Sci Rep, 7(476), 2017.
\newblock \url{https://www.nature.com/articles/s41598-017-00516-6}.

\bibitem{Uus2020}
Uus, A., Zhang, T., Jackson. L. H., Roberts, T. A., Rutherford, M. A., Hajnal, J. V., Deprez, M.
\newblock Deformable Slice-to-Volume Registration for Motion Correction of Fetal Body and Placenta MRI.
\newblock IEEE Trans Med Imaging, 39(9):2750--2759, September 2020.
\newblock DOI: 10.1109/TMI.2020.2972867.
\newblock \url{https://ieeexplore.ieee.org/document/8971472}.

\bibitem{Kuklisova2012}
Kuklisova-Murgasova, M., Quaghebeur, G., Rutherford, M. A., Hajnal, J. V., Schnabel, J. A.
\newblock Reconstruction of fetal brain MRI with intensity matching and complete outlier removal.
\newblock Med Image Anal, 16(8):1550--1564, December 2012.
\newblock DOI: 10.1016/j.media.2012.06.001.
\newblock \url{https://www.sciencedirect.com/science/article/pii/S1361841512000845}.

\bibitem{Uus2023Bounti}
Uus, A. U., Kyriakopoulou, V., Makropoulos, A., Fukami-Gartner, A., Cromb, D., Davidson, A., et al.
\newblock BOUNTI: Brain vOlumetry and aUtomated parcellatioN for 3D feTal MRI.
\newblock bioRxiv, 2023.
\newblock DOI: 10.1101/2023.04.18.537347.
\newblock \url{https://www.biorxiv.org/content/early/2023/04/27/2023.04.18.537347}.

\bibitem{Aviles2023}
Aviles Verdera, J., Story, L., Hall, M., Finck, T., Egloff, A., Seed, P. T., et al.
\newblock Reliability and Feasibility of Low-Field-Strength Fetal MRI at 0.55 T during Pregnancy.
\newblock Radiology, 309(1):e223050, 2023.
\newblock DOI: 10.1148/radiol.223050.
\newblock \url{https://doi.org/10.1148/radiol.223050}.

\bibitem{Ponrartana2023}
Ponrartana, S., Nguyen, H. N., Cui, S. X., Tian, Y., Kumar, P., Wood, J. C., Nayak, K. S.
\newblock Low-field 0.55 T MRI evaluation of the fetus.
\newblock Pediatr. Radiol., March 2023.
\newblock DOI: 10.1007/s00247-023-05151-8.
\newblock \url{https://link.springer.com/article/10.1007/s00247-023-05151-8}.

\bibitem{Payette2023}
Payette, K., Uus, A., Aviles Verdera, J., Avena Zampieri, C., Hall, M., Story, L., et al.
\newblock An automated pipeline for quantitative T2* fetal body MRI and segmentation at low field.
\newblock ArXiv, August 2023.
\newblock \url{https://arxiv.org/abs/2308.02132}.

\bibitem{Gagoski2022}
Gagoski, B., Xu, J., Wighton. P., Tisdall. M. D., Frost, R., Lo, W. C., Golland, A. van der Kouwe, E. Adalsteinsson, P. E. Grant.
\newblock Automated detection and reacquisition of motion-degraded images in fetal HASTE imaging at 3 T.
\newblock Magn Reson Med, 87(4):1914--1922, April 2022.
\newblock DOI: 10.1002/mrm.29095.
\newblock \url{https://onlinelibrary.wiley.com/doi/full/10.1002/mrm.29095}.

\bibitem{Faghihpirayesh2022}
Faghihpirayesh, R., Karimi, D., Erdogmus, D., Gholipour, A.
\newblock Deep Learning Framework for Real-time Fetal Brain Segmentation in MRI.
\newblock arXiv, May 2022.
\newblock \url{https://arxiv.org/abs/2205.01675}.

\bibitem{Salehi2018}
Salehi, S. S. M., Hashemi, S. R., Velasco-Annis, C., Ouaalam, A., Estroff, J. A., Erdogmus, D., et al.
\newblock Real-time automatic fetal brain extraction in fetal MRI by deep learning.
\newblock In: 2018 IEEE 15th International Symposium on Biomedical Imaging (ISBI 2018), pp. 720--724, April 2018.
\newblock DOI: 10.1109/ISBI.2018.8363542.
\newblock \url{https://ieeexplore.ieee.org/document/8363542}.

\bibitem{Singh2020}
Singh, A., Salehi,  S. S. M., Gholipour, A.
\newblock Deep Predictive Motion Tracking in Magnetic Resonance Imaging: Application to Fetal Imaging.
\newblock IEEE Transactions on Medical Imaging, 39(11):3523--3534, 2020.
\newblock DOI: 10.1109/TMI.

\bibitem{Neves_Silva2023}
Neves Silva, S., Aviles Verdera, J., Tomi-Tricot, R., Neji, R., Uus, A., Grigorescu, I., et al.
\newblock Real-time fetal brain tracking for functional fetal MRI,
\newblock Magn. Reson. Med.,
\newblock Jul 2023,

\bibitem{Xue2021-co}
Xue, H., Artico, J., Fontana, M., Moon, J. C., Davies, R. H., Kellman, P.
\newblock Landmark Detection in Cardiac MRI by Using a Convolutional Neural Network,
\newblock Radiol Artif Intell,
\newblock Sep 2021,

\bibitem{Yang2020-nm}
Yang, X., Tang, W. T., Tjio, G., Yeo, S. Y., Su, Y.
\newblock Automatic detection of anatomical landmarks in brain MR scanning using multi-task deep neural networks.
\newblock Neurocomputing, 396:514--521, July 2020.
\newblock DOI: 10.1016/j.neucom.2020.04.078.

\bibitem{Lei2023-wf}
Lei, K., Syed. A. B., Zhu, X., Pauly, J. M., Vasanawala, S. V.
\newblock Automated MRI Field of View Prescription from Region of Interest Prediction by Intra-Stack Attention Neural Network.
\newblock Bioengineering (Basel), 10(1), January 2023.
\newblock DOI: 10.3390/bioengineering10010005.

\bibitem{Hoffmann2021-hd}
Hoffmann, M., Turk, E. A., Gagoski, B., Morgan, L., Wighton, P., Tisdall, M. D., et al.
\newblock Rapid head-pose detection for automated slice prescription of fetal-brain MRI.
\newblock Int. J. Imaging Syst. Technol., 31(3):1136--1154, September 2021.
\newblock DOI: 10.1002/ima.22651.

\bibitem{Xu2019-ls}
Xu, J., Zhang, M., Turk, E. A., Zhang, L., Grant, P. E., Ying, K., et al.
\newblock Fetal Pose Estimation in Volumetric MRI Using a 3D Convolution Neural Network.
\newblock In: Medical Image Computing and Computer Assisted Intervention -- MICCAI 2019, pp. 403--410, 2019.

\bibitem{Hansen2013}
Hansen, M. S., Sørensen, T. S.
\newblock Gadgetron: an open source framework for medical image reconstruction.
\newblock Magn Reson Med, 69(6):1768--1776, June 2013.
\newblock DOI: 10.1002/mrm.24383.

\bibitem{Uus2021UNET}
Uus, A., Grigorescu, I., van Poppel, M., Hughes, E., Steinweg, J., Roberts, T., et al.
\newblock 3D UNet with GAN discriminator for robust localization of the fetal brain and trunk in MRI with partial coverage of the fetal body.
\newblock bioRxiv, 2021.
\newblock DOI: 10.1101/2021.06.23.449574.
\newblock \url{https://www.biorxiv.org/content/early/2021/06/24/2021.06.23.449574}.

\bibitem{Isensee2021}
Isensee, F., Jaeger, P. F., Kohl, S. A. A., Petersen, J., Maier-Hein, K. H.
\newblock nnU-Net: a self-configuring method for deep learning-based biomedical image segmentation.
\newblock Nat. Methods, 18(2):203--211, February 2021.
\newblock DOI: 10.1038/s41592-020-01008-z.

\bibitem{Salehi2019}
Salehi, S. S. M., Khan, S., Erdogmus, D., Gholipour, A.
\newblock Real-Time Deep Pose Estimation With Geodesic Loss for Image-to-Template Rigid Registration.
\newblock IEEE Transactions on Medical Imaging, 38(2):470--481, 2019.
\newblock DOI: 10.1109/TMI.2018.2866442.

\bibitem{She2023-vl}
She, J., Huang, H., Ye, Z., Huang, W., Sun, Y., Liu, C., et al.
\newblock Automatic biometry of fetal brain MRIs using deep and machine learning techniques.
\newblock Sci. Rep., 13(1):17860, October 2023.
\newblock DOI: 10.1038/s41598-023-17202-0.
\end{thebibliography}
\end{document}